\documentstyle[11pt]{article}
\begin{document}
\begin{titlepage}

\title{Grading of Spinor Bundles
and Gravitating Matter in Non-Commutative Geometry}

\author{~\\~\\C.~Klim\v c\'{\i}k$^1$,
 A.~Pompo\v s$^2$ and V.~Sou\v cek$^3$\\
{}~\\~\\ \small $^1$ Theory Division of the Nuclear Centre, Charles
University,\\
\small   V Hole\v sovi\v ck\'ach 2, CS-180 00 Prague 8, Czech Republic\\~\\
\small  $^2$ Department of Theoretical Physics, Charles University,\\
\small   V Hole\v sovi\v ck\'ach 2, CS-180 00 Prague 8, Czech Republic\\~\\
\small  $^3$  Mathematical Institute, Charles University,\\
\small Sokolovsk\'a 83, CS-186 00 Prague 8, Czech Republic\\~}

\date{\small April, 1993}
\maketitle

\begin{abstract}

The gravitating matter  is studied
within
    the framework of the non-commutative geometry. The non-commutative
  Einstein-Hilbert action on the product of a four dimensional
   manifold with a discrete space gives the models of matter fields coupled to
t
he
   standard Einstein gravity. The matter multiplet is encoded in the Dirac
   operator which yields the representation of the algebra of the universal
   forms. The general form of the Dirac operator depends on a choice of the
   grading of the corresponding spinor bundle.
    A choice is given, which leads to the nonlinear vector $\sigma$-
    model coupled to the Einstein gravity.
\end{abstract}

\end{titlepage}

\section{ Introduction.}

Geometrical methods in theoretical physics
are of generally recognized relevance for the  model building. Unravelling of
geometrical structure in a theory usually leads to new insights
and results not achieved previously. Indeed, the  dynamics of nonabelian
gauge theories or of string theory is closely connected with their
geometrical structure \cite{Raja}, \cite{Wgsch}. Apart from the use of
 standard geometrical
methods,
the developments of the last decade in mathematical
physics have led to applications of ideas of the so called
non-commutative geometry proposed by A. Connes.
As an example, we can take  Witten's open string field theory
\cite{Wit} or
the formulation of the standard model using ideas of non-commutative
geometry \cite{Con}, \cite{con1}, \cite{con2}.
The latter application was developed by A. Connes himself and, among other
things, it gave a geometrical meaning to the Higgs field.
The approach of A. Connes is in some sense a small deviation from
the standard commutative geometry, because the notion of a point
retains its sense. Non-commutativity enters when metric properties are
defined.

Recently, A.~H.~Chamseddine, G.~Felder and J.~Fr\"ohlich have introduced
Ein\-stein-Hilbert gravity in the non-commutative geometry framework
\cite{Chff1}.
They considered the space-time as the product of the standard
four-dimensional manifold with Kaluza-Klein-like internal space
consisting of two points. Metric aspects are encoded in the
notion of Dirac operator acting on the direct sum $H$ of two copies
of the space of spinor fields. An important role in such a construction
is played by the grading on the space $H.$
The Dirac operator is required to be odd, hence its general form
clearly depends on a choice of the grading.

In this note, we wish to elaborate on this point. We pick up
a different grading than one discussed in
 Ref. \cite{Chff1} and we consider the most general
Dirac operator consistent with this grading.
While in the ChFF approach, the general Dirac operator depends on
a standard (commutative) vierbein and two scalar fields,
in our case the scalar fields are replaced by one vector field.
Using the non-commutative Einstein-Hilbert action
and imposing the zero (non-commutative) torsion condition,
we get
a nonlinear vector $\sigma$-model coupled to the standard gravity.
The  resulting Lagrangian has a quite unexpected form and
it indicates the richness of types of  models obtainable
from the non-commutative geometry approach.

\section{ Elements of non-commutative geometry.}
Let us consider
a smooth compact four-dimensional spin manifold $Y$ with a fixed
spin structure and the real algebra
$C^{\infty}(Y)$ of smooth functions on it.
 The commutative algebra~$A$
is defined as the algebra of diagonal matrices
\begin{equation}
f=\pmatrix
{f_1&0\cr
 0&f_2}
\end{equation}
with values in
$C^{\infty}(Y).$
Elements of $A$ can clearly be considered
as smooth functions on the product $Y\times Z_2.$

Let $S$ denote the associated spinor bundle  on $Y.$
The Riemannian metric on $Y$ determines the corresponding
Clifford bundle ${\cal C}(T^*Y).$
Its complexification coincides with the bundle ${\rm End}(S),$
sections of ${\cal C}(T^*Y)$ are called real sections of
${\rm End}(S).$
The group $Z_2$ acts by permutation on the bundle $\tilde{S}=S\oplus S,$
equivariant linear operators on sections of $\tilde{S}$ are those
 commuting with the action.

Then we introduce the Dirac $K$-cycle $(H,D,\Gamma),$ where:
\begin{description}
\item{\rm i)}
the Hilbert space
$H$ is the sum of two copies of the Hilbert space of $L_2$-sections
of the spinor bundle $S,$

\item{\rm ii)} the grading  $\Gamma$ is defined by
\begin{equation}
\Gamma=\pmatrix
{\gamma_5& 0\cr
    0& \gamma_5}
\end{equation}

\item{\rm iii)} the Dirac operator $ D$
 is an odd equivariant self-adjoint first order elliptic differential
operator on the space $C^{\infty}(S)\oplus C^{\infty}(S),$

\item{\rm iv)}
For each $f\in A,\; [D,f]$ is multiplication by a real section
of ${\rm End}(\tilde{S}),$ i.e. by a matrix with entries in
real sections of ${\rm End}(S).$
\end{description}

A general scheme of the non-commutative geometry can be applied now
(see \cite{con1}, \cite{con2}, \cite{Chff1}).
The algebra $\Omega^*(A)$ of universal forms over $A$ is generated
by symbols $f\in A$ with degree 0 and $df, f\in A$ with degree $1$
with relations $d(fg)=df\;g + f\;dg, f,g\in A, $ and $d 1=0.$
The representation $\pi$
of $\Omega^*(A)$ given by
$\pi(fdg_1\ldots dg_k)=f[D,g_1]\ldots [D,g_k]$
determines then a two-sided ideal $I={\rm Ker}(\pi)+ d{\rm Ker}(\pi)$
in the algebra $\Omega^*(A)$ and a new graded differential algebra
$\Omega_D^*(A)$ is defined as
the quotient $\Omega^*(A)/I.$

A connection $\nabla:\Omega^1(A)\mapsto \Omega^1(A)\otimes
\Omega^1(A)$ on the module $E=\Omega^1(A)$ over $A$
can be extended uniquely to a linear
map $\nabla:\Omega^p(A)\otimes_A E\mapsto \Omega^{p+1}\otimes_A E$
with the property
\begin{equation}
 \nabla(\alpha\varphi)=d\alpha\,\varphi+(-1)^p\alpha\nabla\varphi,\;
\alpha\in \Omega^p(A),\varphi\in\Omega^*(A)\otimes_A E.
\end{equation}
The curvature and torsion of the connection $\nabla$ are defined by
$R(\nabla)=-\nabla^2$ and $T(\nabla)=d-m\circ\nabla,$
where $m:\Omega^1(A)\otimes\Omega^1(A)\mapsto
\Omega^2(A)$ denotes multiplication in $\Omega^*(A).$

In a local situation, let $E^A,A=1,\ldots,N,$ be an orthonormal
basis of $\Omega^1_D(A)$ with respect to the Riemannian metric given
by
\begin{equation}
G(\alpha,\beta)={\rm tr}(\pi(\alpha)^*\pi(\beta)),\;
\alpha,\beta\in\Omega^1_D(A).
\end{equation}
The coordinate descriptions of the connection matrix, torsion and
curvature are defined by
\begin{equation}
\pi(\nabla)E^A=-\sum_B\Omega^{AB}\otimes E^B,
\end{equation}
\begin{equation}
\pi(T(\nabla))E^A=T^A,
\end{equation}
\begin{equation}
\pi(R(\nabla))E^A=\sum_B R^{AB}\otimes E^B
\end{equation}
and computed by
\begin{equation}T^A=\pi(d\tilde{E}^A)+\sum_B\Omega^{AB}E^B\end{equation}
and
\begin{equation}R^{AB}=\pi(d\tilde{\Omega}^{AB})+\sum_C\Omega^{AC}\Omega^{CB},\e
nd{equation}
where
$\tilde{E}^A$ and $\tilde{\Omega}^{AB}$ are representatives of
$E^A,$ resp. $\Omega^{AB},$ in $\Omega^1(A).$

\section{ The Einstein-Hilbert action.}
Let us introduce the inner product on two-forms defined through the
identification of $\Omega^2_D(A)$ with $\pi(\Omega^2(A))\cap
\pi(dKer(\pi|_{\Omega^1(A)})^{\perp})$ as follows
\begin{equation}
(\alpha,\beta)=Tr_{\omega}(\pi(\alpha)^*\pi(\beta)|D|^{-4})
\end{equation}
where $Tr_{\omega}$ is the Dixmier trace \cite{Chff1}.
 As a gravity action, we take the generalized Einstein-Hilbert action
 \cite{Chff1} given by means of the Dixmier trace i.e.
 \begin{equation}
 Tr_{\omega}((E^AE^B)^*(R^{AB})|D|^{-4})
 \end{equation}
 where ${E^A}$ is the orthonormal basis in $\Omega^1_D(A)$ and $R^{AB}$ the
 curvature two-form.The representatives of the two forms are taken to be
 orthogonal to \newline  $dKer (\pi|_{\Omega^1(A)})$.
 This action reduces \cite{Chff1} to
 the integral over Y
 \begin{equation}
 S=\int_{Y}tr((E^AE^B)^*R^{AB})\sqrt{g}d^4y
 \end{equation}
 where the trace is over $End(S_{y}\oplus S_{y})$.
 In order to work out the action in components, we have to know the
 concrete expressions for the two forms needed. We shall proceed as
 follows.
  First we write down the most general form of the Dirac operator D,
  compatible with self-adjointness, reality and oddness (with respect to the
  grading (2)). It reads

\begin{equation}
D=\pmatrix{ \gamma^ae^{\mu}_a\partial_{\mu}+... & \gamma^{\mu}\gamma^{5}V_{\mu}
\cr \gamma^{\mu}\gamma^{5}V_{\mu} & \gamma^ae^{\mu}_a \partial _{\mu}+... }
\end{equation}

Such ansatz reflects  also the fact that $[D,f]$ should be a multiplication
operator for $f\in A$.
 Here $e^{\mu}_a$ and $A_{\mu}$ are real functions, the dots are zero order
 contributions which do not contribute to $\pi$. The ellipticity of D
 implies that $e^{\mu}_a\partial_{\mu} $ is a basis of the tangent space.
 Recall, that $\gamma^a\gamma^b+\gamma^b\gamma^a=
 -2\delta^{ab}, \gamma^5\equiv\gamma^1\gamma^2\gamma^3\gamma^4$ and
 $(\gamma^a)^*=-\gamma^a.$
  Consider now the universal one-form $\alpha=\sum_ia_idb_i \in
  \Omega^1(A)$. It is not difficult to find  its representative
\begin{equation}
\pi(\alpha)=\sum_ia_i[D,b_i]=\pmatrix{\gamma^{\mu}\alpha_{1\mu} &
V_{\mu}\gamma^{\mu}\gamma^5\alpha_5 \cr
-V_{\mu}\gamma^{\mu}\gamma^5\tilde\alpha_5 & \gamma^{\mu}\alpha_{2\mu} }
\end{equation}  where \begin{equation}
\alpha_{1\mu}=\sum_ia_{i1}\partial_{\mu}b_{i1} ,\quad
\alpha_{2\mu}=\sum_ia_{i2}\partial_{\mu}b_{i2}
\end{equation} \begin{equation}
\alpha_5=\sum_ia_{i1}(b_{i2}-b_{i1}),\quad
\tilde\alpha_5=\sum_ia_{i2}(b_{i2}-b_{i1})
\end{equation}
 The Riemannian metric G: $\Omega^1_D(A)\otimes\Omega^1_D(A)\to A$
 can be expressed by   (4)
 \begin{equation}
 G(\alpha,\beta)=tr(\pi(\alpha)^*\pi(\beta))
 \end{equation}
 hence
 \begin{equation}
 G(\alpha,\beta)=(g^{\mu\nu}\alpha_{1\mu}\beta_{1\nu}+
 g^{55}\tilde\alpha_5\tilde\beta_5)\oplus(g^{\mu\nu}\alpha_{2\mu}\beta_{2\nu}
 +g^{55}\alpha_5\beta_5)
 \end{equation}
 where
  \begin{equation}
 g^{\mu\nu}=-tr(\gamma^{\mu}\gamma^{\nu})=e^{\mu}_ae^{\nu}_a  \end{equation}
   \begin{equation}
 g^{55}=g^{\mu\nu}V_{\mu}V_{\nu}\equiv V^2
 \end{equation}
  Knowing the metric, we may introduce local orthonormal bases {$E^A$} in
  $\Omega^1_D(A)$. As in \cite{Chff1}, we use the capital letters A,B... to
denote  indices taking the values 1 to 5, and lower case letters a,b,... taking
  the values 1 to 4. The basis is

\begin{equation}
E^{a}=\pmatrix{ \gamma^{a} & 0 \cr 0 & \gamma^{a} }=\pmatrix{
\gamma^{\mu}e^{a}_
{\mu}
& 0 \cr 0 & \gamma^{\mu}e^{a}_{\nu} }
\end{equation}
\begin{equation}
E^{5}=\pmatrix{ 0 &\frac{V_{\mu}\gamma^{\mu}}{\sqrt{V^2}} \gamma^{5} \cr
-\frac{V_{\mu}\gamma^{\mu}}{\sqrt{V^2}} \gamma^{5} & 0 }
\end{equation}
 Now we have to understand two-forms $\Omega^2_D(A)$. First we
 identify the space of the "auxiliary"fields $\pi(dKer(\pi|_{\Omega^1(A)}))$
 \cite{Chff1}. Clearly, from \\$\pi(d\alpha)=\sum_i[D,a_i][D,b_i]$
 we obtain for $\alpha\in Ker(\pi)$
 \begin{equation}
 \pi(d\alpha)=\pmatrix{ -g^{\mu\nu}\partial_{\mu}a_{i1}\partial_{\nu}b_{i1}
 & -2\gamma^{\mu\nu}\gamma^5V_{\nu}a_{i1}\partial_{\mu}b_{i2} \cr
 2\gamma^{\mu\nu}\gamma^5V_{\nu}a_{i2}\partial_{\mu}b_{i1} &
 -g^{\mu\nu}\partial_{\mu}a_{i2}\partial_{\nu}b_{i2} }
 \end{equation}
 Clearly, whatever expression of the form
 \begin{equation}
 \pmatrix{ X_1 & \gamma^{\mu\nu}\gamma^5V_{\nu}Y_{\mu} \cr
 \gamma^{\mu\nu}\gamma^5V_{\nu}\tilde Y_{\mu} & X_2 }
 \end{equation}
 can be obtained in this way, where $X_1$,$X_2$,$Y_{\mu}$ and $\tilde
 Y_{\mu}$ are given.
  If we consider an arbitrary one-form $\alpha$, then the representative
  $\pi(d\alpha)$ orthogonal to all auxiliary fields (24) is uniquely given
  by
  \begin{equation}
  \pi(d\alpha)=\pmatrix { \gamma^{\mu\nu}\partial_{\mu}\alpha_{1\nu}&
 -g^{\mu\nu}\gamma^5V_{\nu}(\partial_{\mu}\alpha_5+\alpha_{1\mu}-\alpha_{2\mu})
 \cr g^{\mu\nu}\gamma^5V_{\nu}(\partial_{\mu}\tilde
 \alpha_5+\alpha_{1\mu}-\alpha_{2\mu})
 & \gamma^{\mu\nu}\partial_{\mu}\alpha_{2\nu} }
  \end{equation}
    The orthogonality is understood with respect to the inner product on
    $\Omega^2(A)$ defined     by the Dixmier trace (10).
        Now consider the connection $\nabla$ unitary with respect to the given
    K-cycle \cite{Chff1}. The components of the one-form $\pi(\nabla)$ are
denot
ed by
  \begin{equation}
  \Omega^{AB}=\pmatrix{ \gamma^{\mu}\omega^{AB}_{1\mu} & V_{\mu}\gamma^{\mu}
\ga
mma^{5}l^{AB}
\cr -V_{\mu}\gamma^{\mu} \gamma^{5} \tilde l^{AB} &
\gamma^{\mu}\omega^{AB}_{2\m
u} }
  \end{equation}
  The unitarity conditions $(\Omega^{AB})^*=\Omega^{BA}$ implies the
  relations
        \begin{equation}
            \omega^{AB}_{1\mu} =-\omega^{BA}_{1\mu}  ,\omega^{AB}_{2\mu}
           =-\omega^{BA}_{2\mu},
            \tilde l^{AB}=- l^{BA}
            \end{equation}

 The torsion one-form and the curvature two-form can be calculated from the
 Cartan structure equations  (8),(9). They are
 \begin{equation}
T^{a}=\pmatrix{ \gamma^{\mu\nu}(\partial_{\mu}e^{a}_{\nu}+\omega^{ab}_{1\mu}
e^{b}_{\nu})
&
g^{\mu\nu}\gamma^{5}V_{\nu}(l^{ab}e^{b}_{\mu}-
\frac{\omega^{a5}_{1\mu}}{\sqrt{V^2}})
 \cr
-g^{\mu\nu}\gamma^{5}V_{\nu}(\tilde
l^{ab}e^b_{\mu}-\frac{\omega^{a5}_{2\mu}}{\sqrt{V^2}}) &
\gamma^{\mu\nu}(\partial_{\mu}e^{a}_{\nu}+\omega^{ab}_{2\mu}e^{b}_{\nu}) }
\end{equation}
\vspace{4mm}
\begin{equation}
T^{5}=\pmatrix{ \gamma^{\mu\nu}\omega^{5b}_{1\mu}e^{b}_{\nu} & g^{\mu\nu}
\gamma^{5}V_{\nu}(e^{b}_{\mu}l^{5b}-\partial_{\mu}(\frac{1}{\sqrt{V^2}})) \cr
-g^{\mu\nu}\gamma^{5}V_{\nu}(e^{b}_{\mu} \tilde l^{5b}-
\partial_{\mu}(\frac{1}{\sqrt{V^2}}))
& \gamma^{\mu\nu}\omega^{5b}_{2\mu}e^{b}_{\nu} }
\end{equation}
\vspace{4mm}
\begin{equation}
R^{AB}=\pmatrix{ \frac{1}{2}\gamma^{\mu\nu}R^{AB}_{1\mu\nu} & -g^{\mu\nu}
\gamma^{5}V_{\nu}H^{AB}_{\mu} \cr
g^{\mu\nu}\gamma^{5}V_{\nu} \tilde H^{AB}_{\mu} & \frac{1}{2}\gamma^{\mu\nu}
R^{AB}_{2\mu\nu} }
\end{equation}

where
  \begin{equation} R^{AB}_{i\mu\nu}= \partial_{\mu}\omega^{AB}_{i\nu}-
     \partial_{\nu}\omega^{AB}_{i\mu}+
      \omega^{AC}_{i\mu} \omega^{CB}_{i\nu}-
      \omega^{AC}_{i\nu}\omega^{CB}_{i\mu}
  \end{equation}
  \begin{equation}
  H^{AB}_{\mu}=\partial_{\mu} l^{AB}+ \omega^{AB}_{1\mu}  -
  \omega^{AB}_{2\mu} +   l^{CB} \omega^{AC}_{1\mu}-  l^{AC} \omega^{CB}_{2\mu}
  \end{equation}
  \begin{equation}
 \tilde H^{AB}_{\mu}=\partial_{\mu}\tilde l^{AB} +  \omega^{AB}_{1\mu} -
  \omega^{AB}_{2\mu} - \tilde l^{AC} \omega^{CB}_{1\mu} +\tilde l^{CB}
  \omega^{AC}_{2\mu}
  \end{equation}
 The Einstein-Hilbert action (13) therefore reads
 \begin{equation}
 I=\int_Y d^4y\sqrt g[e^{\mu}_ae^{\nu}_b(R^{ab}_{1\mu\nu}+ R^{ab}_{2\mu\nu})+

g^{\mu\nu}g^{\alpha\beta}e^a_{\mu}\frac{V_{\nu}V_{\alpha}}{\sqrt{V^2}}(H^{a5}_{
\beta}+
 \tilde H^{a5}_{\beta}-H^{5a}_{\beta}-\tilde H^{5a}_{\beta})]
 \end{equation}
  In what follows, we shall  restrict our attention to the torsion-free
  theory.
    Requiring  $T^A=0$, we obtain
  \begin{description}
  \item{1)}
  \begin{equation}\qquad\qquad\omega^{ab}_{\mu}\equiv \omega^{ab}_{1\mu}=
\omega
^{ab}_{2\mu}
  ={\rm Levi-Civita}\end{equation}
  \item{2)}

\begin{equation}\qquad\qquad\partial_{\mu}(\frac{1}{\sqrt{V^2}})=e^{b}_{\mu}l^
{5b},l^{5b}=-l^{b5},l^{ab}=l^{ba}\end{equation}
  \item{3)}
  \begin{equation}\sqrt{V^2}~l^{ab}e^b_{\nu}=\omega^{a5}_{1\nu}=-
\omega^{a5}_{2
\nu},\end{equation}
    \begin{equation}\sqrt{V^2}~\tilde
l^{ab}e^b_{\nu}=\omega^{a5}_{2\nu}\end{equ
ation}
  \end {description}
  Inserting the conditions (35),(36),(37),(38) into (34) we get the
 action
  \begin{equation}
  S=\int_Y

d^4y\sqrt{g}[2R+4D_{\beta}(\frac{V^{\mu}V^{\beta}}{\sqrt{V^2}})D_{\mu}(\frac{1
}{\sqrt{V^2}})
  +4V^aV^b(l^{cb}l^{ca}+l^{ab}l^{55}+2l^{ab})]
  \end{equation}
   The fields $l^{ab}$ and $l^{55}$ decouple and we finaly obtain the
   action of the gravitating vector $\sigma$-model
   \begin{equation}
   S=\int_y
   d^4y\sqrt{g}[2R+
Q^{\alpha\beta\gamma\delta}(V)D_{\alpha}V_{\beta}D_{\gamma}V_{\delta}]
   \end{equation}
   where $$
   Q^{\alpha\beta\gamma\delta}(V)=4\frac{1}{(V^2)^3}\Big(V^{\alpha}
V^{\beta}V^{\gamma}V^{\delta}- g^{\alpha\beta}V^{\gamma}V^{\delta}V^2
- g^{\gamma\beta}V^{\alpha}V^{\delta}V^2\Big)
   $$

   Let us conclude with few remarks. The choice of the grading of spinor
    bundle in the work \cite{Chff1} led to the minimally coupled
     masless scalar field.
     This scalar
    field was interpreted as measuring the distance between the points in the
    internal space. In our case such interpretation is not possible. Using
    the Eq.(20), we might say that square $V^2$ measures the distance.
    However, it seems quite clear that in more complicated cases, the
    interpretation of the fields presented in the Dirac operator will
    not be
    straightforward. We should add also another remark concerning
    the concrete solutions of the non-commutative models.
     In the case of minimally coupled scalar field the classification
    of the static spherically symmetric asymptotically flat
     solutions is known \cite{xant}.
    In the case of the vector $\sigma$-model such analysis is absent and is
     the matter for future research \cite{KliPo}.

\vskip 2pc

    \leftline{\bf Acknowledgement.}

\vskip 1pc

   \noindent We thank J.~Bure\v s and R.~Tom\'a\v sek for discussions.


\begin{thebibliography}{19}
 \bibitem{Raja}{R.~Rajaraman, {\it Solitons and Instantons}, North-Holland
 P.C., Amsterdam (1982)}
  \bibitem{Wgsch}{M.~Green, J.~Schwarz and E.~Witten, {\it Superstring
 Theory}, Cambridge Press, Cambridge (1987)}.
  \bibitem{Wit}{E.~Witten, {\it Nucl.Phys.}, {\bf B268}, 253 (1986).}
  \bibitem{Con}{ A.~Connes, in {\it The interface of mathematics and
   particle physics}, Clarendon press, Oxford 1990, Eds D.~Quillen, G.~Segal
    and S.~Tsou}
  \bibitem{con1}{ A.~Connes and J.~Lott, {\it Nuclear Phys. B Proc. Supp.}
  {\bf 18B} 29 (1990), North-Holland, Amsterdam.}
  \bibitem{con2}{ A.~Connes and J.~Lott, {\it The metric aspect of
  noncommutative geometry}, appear in Proceedings of the 1991
  Carg\`ese summer school.}
    \bibitem{Chff1}{ A.~H.~Chamseddine, G.~Felder, J.~Fr\"ohlich, {\it Gravity
     in Non-Commutative
        Geometry}, Z\"urich preprint 1992, {\bf ETH-TH/1992-18},  }
  \bibitem{xant}{B.~C.~Xanthopoulos and T.~Zannias, {\it Phys.
  Rev.}, {\bf D40}, 2564 (1989).}
 \bibitem{KliPo}{C.~Klim\v c\' {\i}k and A.~Pompo\v s, in preparation}
\end{thebibliography}
\end{document}